\documentclass[twocolumn,
superscriptaddress,
 amsmath,amssymb,
 aps,
prl,
]{revtex4-1}

\usepackage{graphicx}% Include figure files
\usepackage{dcolumn}% Align table columns on decimal point
\usepackage{bm}% bold math

\begin{document}

%\title{Indistinguishable photons from dissipative emitters in cavities}
%\title{Restoring photon indistinguishability from uncoherent emitters by cavity funneling}
%\title{Proposed efficient production of indistinguishable photons from strongly dissipative emitters by cavity funneling}

%\title{Cavity-funneled emission of indistinguishable photons from strongly dissipative systems}
\title{Cavity-funneled generation of indistinguishable single photons from strongly dissipative quantum emitters}

\author{Thomas Grange}
\affiliation{Universit\'e Grenoble-Alpes, ``Nanophysics and Semiconductors'' joint team, 38000 Grenoble, France}
\affiliation{CNRS, Institut N\'{e}el, ``Nanophysics and Semiconductors'' joint team, 38000 Grenoble, France}

\author{ Gaston Hornecker}
\affiliation{Universit\'e Grenoble-Alpes, ``Nanophysics and Semiconductors'' joint team, 38000 Grenoble, France}
\affiliation{CNRS, Institut N\'{e}el, ``Nanophysics and Semiconductors'' joint team, 38000 Grenoble, France}
\affiliation{CEA, INAC-SP2M, ``Nanophysics and Semiconductors'' joint team, 38000 Grenoble, France}

\author{David Hunger}
\affiliation{Fakult\"{a}t f\"ur Physik, Ludwig-Maximilians-Universit\"at, Schellingstr. 4, 80799 M\"unchen, Germany}
 
  \author{Jean-Philippe Poizat}
\affiliation{Universit\'e Grenoble-Alpes, ``Nanophysics and Semiconductors'' joint team, 38000 Grenoble, France}
\affiliation{CNRS, Institut N\'{e}el, ``Nanophysics and Semiconductors'' joint team, 38000 Grenoble, France}
   
   \author{Jean-Michel G\'erard}
 \affiliation{Universit\'e Grenoble-Alpes, ``Nanophysics and Semiconductors'' joint team, 38000 Grenoble, France}
\affiliation{CEA, INAC-SP2M, ``Nanophysics and Semiconductors'' joint team, 38000 Grenoble, France}

 \author{Pascale Senellart}
 \affiliation{CNRS, Laboratoire de Photonique et de Nanostructures,  91460 Marcoussis, France}
  
 \author{Alexia Auff\`{e}ves}
\affiliation{Universit\'e Grenoble-Alpes, ``Nanophysics and Semiconductors'' joint team, 38000 Grenoble, France}
\affiliation{CNRS, Institut N\'{e}el, ``Nanophysics and Semiconductors'' joint team, 38000 Grenoble, France}

\date{\today}

\begin{abstract}
We investigate theoretically
the generation of indistinguishable single photons from a strongly dissipative quantum system placed inside an optical cavity.
The degree of indistinguishability of photons emitted by the cavity
%as well as the efficiency of the process
%Efficiency of photon emission by the cavity and indistinguishability are
is calculated as a function of the emitter-cavity coupling strength and the cavity linewidth.
%For a Markovian system with pure dephasing rates much larger than its free-space radiative emission rate
For a quantum emitter subject to strong pure dephasing, our calculations reveal that an unconventional regime %combining
%almost-perfect
of high
indistinguishability
%(cavity-funneled)
%efficiency beyond spectral filtering
%good emission efficiency
can be reached for moderate emitter-cavity coupling strengths and high quality factor cavities.
In this regime, the broad spectrum of the dissipative quantum system is funneled into the narrow lineshape of the cavity.
The associated efficiency is found to greatly surpass spectral filtering effects.
%For room-temperature solid-state emitters, this regime is found to be experimentally-accessible parameters of moderate James-Cuming coupling strength and low cavity linewidth.
Our findings open the path towards
%the efficient generation of indistinguishable photons by independent solid-state quantum emitters at room temperature.
%the realization of solid-state indistinguishable single-photon emitting devices working at room temperature.
%all-solid-state generation of indistinguishable photons at room temperature.
on-chip scalable indistinguishable-photon emitting devices operating at room temperature.

% tin the space of James-Cuming coupling strength-cavity linewidth
%We derive a simple expression (efficient calculation) for the indistinguishability that allows us to map the degree of %indistinguishability of the photons emitted in the cavity mode as a function of the system parameters. 
%For a given pure dephasing rate, we show that near-perfect indistinguishability are possible in two opposite regimes. The %physics and the experimental/technological relevance of the different regimes are discussed.
%Pure dephasing limits the degree of indistinguishability
\end{abstract}

\maketitle

Indistinguishable single photons are the building blocks of various optically-based quantum information applications such as linear optical quantum computing \cite{knill2001scheme,kok2007linear}, boson sampling \cite{spring2013boson, broome2013photonic,crespi2013integrated,tillmann2013experimental,spagnolo2014experimental}, quantum teleportation \cite{fattal2004quantum} or quantum networks \cite{kimble2008quantum}.
Indistinguishable photons are usually generated either using 
%nonlinear effects from classical electromagnetic waves
parametric down conversion \cite{eisaman2011invited}, or alternatively directly from a single two-level quantum emitter such as atoms, color centers, quantum dots or organic molecules \cite{santori2002indistinguishable,kiraz2005indistinguishable,patel2010two,lettow2010quantum,gazzano2013bright,he2013demand,muller2014demand,monniello2014indistinguishable,sipahigil2014indistinguishable,wei2014deterministic}. 
%Non-linear processes combined with heralding detection
Parametric down conversion
is presently the most mature technology available, 
%(and possible multi-photon generation)
%but the intrinsic nondeterministic emission of photons combined with the very low efficiency of the nonlinear processes make it hardly on-chip scalable.
but the usual low efficiency of the nonlinear processes
%together with the thermal distribution of the emitted photons are
is a severe limitation to the scalability of such sources.
On the other hand, sources based on single solid-state quantum systems have been greatly developped in the last decade, as they hold the promise to combine indistinguishable, on-demand, energy-efficient, electrically drivable and scalable characteristics.
% in addition to being a photonic--solid-state interface. 
%A prerequiste for the implementation of optical quantum computing schemes \cite{} is the production of
%A robust implementation of indistinguishable single-photon emitters is highly desirable for quantum information applications such as quantum teleportation \cite{fattal2004quantum} or linear optical quantum computing schemes \cite{knill2001scheme,kok2007linear}. 
However, except at cryogenics temperature, solid-state systems emitting single-photons are subject to strong pure dephasing processes \cite{borri01,bayer2002temperature,berthelot2006unconventional,kako2006gallium,grange2009decoherence,bounouar2012ultrafast,albrecht2013coupling,albrecht2014narrow,rogers2014multiple}, making them at first view inappropriate for quantum applications requiring photon indistinguishability.
%Indistinguishable single photons are the building blocks of optical quantum computing protocols \cite{knill2001scheme,kok2007linear}.

%The indistinguishability of photons can be tested in a Hong-Ou-Mandel (HOM) interferometer \cite{hong1987measurement}: if two perfectly indistinguishable photon wave packets are sent on the two input ports of a balanced beam splitter, they coalesce on one of the two output ports and no coincidence detection is measured due to a two-photon destructive interference effect. For pulsed single-photon wave-packets entering an HOM interferometer, the probability of coincidence events $p_{\text{c}}$ can be measured. The degree of indistinguishability, or mean wave-packet overlap, can be defined as
%\begin{equation}
%I = 1 - 2p_{\text{c}},
%\label{defI}
%\end{equation}
%and should be unity (resp. zero) for a coherent (resp. incoherent) single-photon wavepacket.

A two-level quantum emitter (QE) coupled only to
vacuum fluctuations should emit perfectly indistinguishable photons. However, as soon as pure dephasing of the QE occurs, the degree of indistinguishability of the emitted photons is reduced to \cite{bylander2003interference}
%When only population decay of an excited two-level system leads to perfectly indistinguishable photons, additionnal pure dephasing processes affecting the TLS decreases the degree of indistinguishability to
%For a Markovian two-level system (TLS) emitting photons, pure-dephasing processes limit the degree of indistinguishability to 
\begin{equation}
I = \frac{\gamma}{\gamma + \gamma^*}
%= \frac{1/T_1}{1/T_1 + 1/2T_2^*}
%= \frac{2T_2^*}{2T_2^* + T_1}
= \frac{T_2}{2T_1},
\label{ITLS}
\end{equation}
where $\gamma=1/T_1$ is the population decay rate, $\gamma^*/2 = 1/T_2^*$ the pure dephasing rate, and $1/T_2 = 2/T_1 + 1/T_2*$ the total dephasing rate.
For solid-state QE emitting photons at room temperature such as color centers, quantum dots or organic molecules, pure dephasing rates are typically several orders of magnitude larger than the population decay rate (typically ranging from 3 to 6 orders of magnitude) \cite{borri01,bayer2002temperature,kako2006gallium,bounouar2012ultrafast,albrecht2013coupling,albrecht2014narrow,rogers2014multiple,lounis2000single,kiraz2005indistinguishable,lettow2010quantum}. Hence the intrinsic indistinguishability given by Eq.~\ref{ITLS} is almost zero.
A possible way to enhance the indistinguishability is to spectrally filter the emitted photons a posteriori. However, this linear-filter strategy leads to a very low efficiency.
Engineering of both efficiency and indistinguishability are possible by placing the dissipative QE in an optical cavity
\cite{gerard1998enhanced,andreani1999strong,moreau2001single,reithmaier2004strong,varoutsis2005restoration,hennessy2007quantum,di2012controlling,albrecht2013coupling,
kaupp2013scaling,gazzano2013bright,riedrich2014deterministic,pathak2010coherently,close2012overcoming,kaer2013microscopic,kaer2014decoherence}.
A usual strategy is then to use the Purcell effect to enhance the spontaneous emission, as in Eq.~\ref{ITLS} an increase in $\gamma$ results in an increase of $I$.
However, reaching Purcell factors larger than $\gamma^*/\gamma$ for room-temperature solid-state systems appears to be well beyond the present experimental state of the art.

In this letter we propose a realistic and robust way to generate
%efficiently
highly indistinguishable photons from strongly dissipative QE (i.e. for $\gamma^* \gg \gamma$).
The idea is to exploit a cavity-quantum-electrodynamics (cavity-QED) regime of low cavity linewidth and moderate cavity-emitter coupling, in which the broad spectrum of the dissipative QE is funneled into the narrow emission line of the cavity.
In this regime, high indistinguishability is predicted together with efficiencies orders of magnitude higher than spectral filtering.
Insights into the full quantum calculation are gained by semiclassical derivations of indistinguishability in limiting cases of dissipative cavity QED.
%In addition, an heralding scheme of the emitted photons based on a $\Lambda$ system is proposed.
%In addition to a fully-quantum calculation of indistinguishability, analytic formulae are derived to gain new insights into different limiting cases of strongly dissipative cavity quantum electrodynamics (cavity QED).
%In addition, new insights into strongly dissipative cavity quantum electrodynamics (cavity QED) are gained by comparing the full quantum calculation of indistinguishability to analytic formulae in limiting cases.

As depicted in Fig.\ref{schema}, we consider a two-level QE system $\{|\psi_g\rangle,|\psi_e\rangle\}$ coupled to a cavity mode whose Fock states are denoted $\{|0\rangle,|1\rangle,...\}$.
%, which is modelled through the Jaynes-Cummings Hamiltonian \cite{}.
All the dissipative terms are assumed to be described within the Markov approximation \cite{auffeves2009pure,auffeves2010controlling}.
%Note that non-Markovian dephasing effects, in particular phonon sidebands in solid-state artificial atoms, are outside the scope of this work \cite{}.
The relevant parameters are: the QE decay rate $\gamma$ (which may include radiative as well as non-radiative components), the cavity decay rate $\kappa$, the pure dephasing rate $\gamma^*$ ;  $g$ is the dipolar coupling between the QE and the cavity mode (see Fig.\ref{schema}). The emitter-cavity detuning is set to zero (i.e. perfect resonance). 
For simplicity, we assume an instantaneous excitation of the QE, so that only one quantum of excitation can be transferred to the cavity. 
Within the rotating-wave approximation, it is therefore sufficient to investigate the dissipative quantum dynamics in the two-dimensional Hilbert space formed by $\{|\psi_e,0\rangle,|\psi_g,1\rangle\}$. 
\begin{figure}
\begin{centering}
\includegraphics[width=0.45\textwidth]{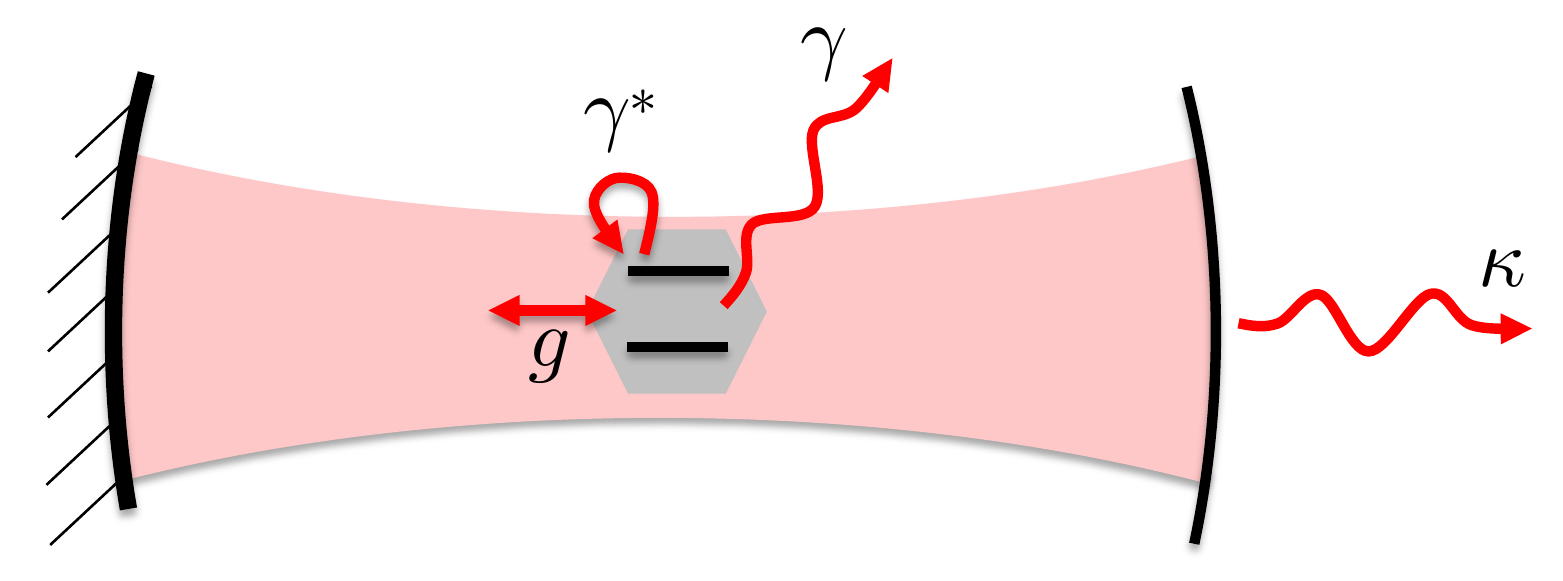}
\par\end{centering}
\caption{Schematic of the system under study: a dissipative two-level emitter coupled to an optical-cavity mode.}
\label{schema}
\end{figure}
%The indistinguishability of photons is typically tested in a Hong-Ou-Mandel (HOM) interferometer \cite{hong1987measurement}: if two perfectly indistinguishable photon wave packets are sent on the two input ports of a balanced beam splitter, they coalesce on one of the two output ports and no coincidence detection is measured due to a two-photon destructive interference effect. For pulsed single-photon wave-packets entering an HOM interferometer, the probability of coincidence events $p_{\text{c}}$ can be measured.
%The degree of indistinguishability, or mean wave-packet overlap, can be defined as
%\begin{equation}
%I = 1 - 2p_{\text{c}},
%\label{defI}
%\end{equation}
%and should be unity (resp. zero) for a coherent (resp. incoherent) single-photon wavepacket.
The degree of indistinguishability of photons can be defined by the probability of two-photon interference in a Hong-Ou-Mandel experiment \cite{hong1987measurement}.
For a single-photon emitter, this indistinguishability figure of merit
%defined in Eq.~\ref{defI}
reads \cite{bylander2003interference,kiraz2004quantum}:
\begin{equation}
I =  \frac{\int_{0}^{\infty} dt \int_0^{\infty} d\tau \vert\langle \hat{a}^{\dagger}(t+\tau)\hat{a}(t)\rangle\vert^2}{\int_{0}^{\infty} dt \int_0^{\infty} d \tau \langle \hat{a}^{\dagger}(t)\hat{a}(t)\rangle\langle \hat{a}^{\dagger}(t+\tau)\hat{a}(t+\tau)\rangle}
\label{Iformula}
\end{equation}
where $a^{(\dagger)}$ are the ladder operators of the EM mode in which the photons are emitted.
%Hence perfect indistinguishability requires the square modulus of the field correlation function to follows exactly the intensity correlations.
%This condition can divided into two: (i) phase correlation has to be maintened during cavity feeding, and (ii) the cavity mode has to decay in a coherent way.
This equation imposes the necessary condition for perfectly indistinguishable photons that time correlations of the EM field decay the same way as the intensity, i.e. that photons are Fourier-transform limited.
%In addition, it also imposes that the population transfer from the QE to the cavity, if incoherent, occurs on a shorter time scale than the cavity decay. We will refer to the former as the time-jitter effect, in an analogous way as the one occuring when the QE is incoherently excited from a high energy state \cite{kiraz2004quantum}.
%The feeding of the cavity must be either coherent or incoherent but
%For the assumed Markovian dissipation, 
The calculation of the above quantities can be separated into two steps (see the Supplemental Material \cite{supplemental}).
First, we calculate the evolution of the density matrix $\hat{\rho}(t)$
by solving the Lindblad equation.
%\begin{equation}
%\frac{ \partial \rho}{\partial t} = i[\rho,H] + \mathcal{L_{\text{damp}}}
%\end{equation}
\begin{equation}
\hat{\rho}(t) = e^{-i \hat{\mathcal{L}}t} |\psi_e,0\rangle \langle \psi_e,0 |
\end{equation}
where $\hat{\mathcal{L}}$ is the total Liouvillian of the system \cite{supplemental}. % including the dissipative Lindbladt terms.
%and where we assume as initial condition that the QE is excited at $t=0$.
Secondly, we calculate the retarded Green's function,
%, which contains all the spectral information \cite{}
%In order to calculate the correlators, we begin by calculating the retarded Green's function of the system
which reads in the $\{|\psi_e,0\rangle,|\psi_g,1\rangle\}$ basis:
\begin{equation}
\hat{G}^{R}(\omega) = \begin{pmatrix}
  \omega + i\gamma/2 + i\gamma^*/2 & g \\
  g & \omega -\delta + i\kappa/2
 \end{pmatrix}^{-1}.
 \label{retardedGF}
\end{equation}
%Note that the spectrum emitted by the cavity is proportional to the imaginary part of $\langle g,1 | G_{r}(\omega) |g,1\rangle$.
%In general, the equations of motion for the correlators (lesser Green's function) are given by the Kadanoff-Baym equations \cite{}.
%For Markovian systems, the equations are separable, and the so-called quantum regression theorem applies:
%\begin{equation}
%G^<(t+\tau,t) = G^R(\tau) \rho (t)
%\end{equation}
%According to the quantum regression theorem \cite{}
The two-time correlator of the cavity field can be expressed as a product of the retarded propagator
$\hat{G}^{R}(\tau) = -i \int d \omega e^{-i\omega t} \hat{G}^R(\omega)$
and the density matrix \cite{supplemental}:
\begin{equation}
\langle a^{\dagger}(t+\tau)a(t)\rangle\ = \langle \psi_g,1 |  \hat{G}^R(\tau) \hat{\rho} (t) |\psi_g,1\rangle .
\end{equation}

%In this regime, the efficiency can surpass by several order of magnitudes the one obtained from simple spectral filtering.

%The price to pay is a smaller
%The idea is to insert the emitters into cavities with high quality factor
 %uses an EM cavity in a regime where the coherence time of the emitted photons correponds to its lifetime in the cavity mode.

%For a Markovian two-level system, the degree of indistinguishability of the emitted photons is limited by the ratio of pure dephasing rates and decay rates.

In Fig.~\ref{Imap}, we report calculations for strongly dissipative emitters verifying $\gamma^*= 10^4 \gamma $.  This is a typical value for a solid-state QE at room temperature,
%\footnote{this ratio can be increased when the QE is placed in a small mode-volume cavity due to possible decrease of the density of states of resonant free photons}.
and the results are qualitatively similar for any strongly dissipative emitters verifying $\gamma^* \gg \gamma$.
%QE parameters corresponding to the case of silicon vacancy (SiV) centers in diamonds at room temperature \cite{}.
%The results are qualitatively general to any strongly dissipative emitters verifying $\gamma^* \gg \gamma$.
%We take $\gamma$ = 0.16 GHz and $\gamma^*$ = 550 GHz \cite{}.
Without any cavity, the degree of indistinguishability would then amount to $10^{-4}$ according to Eq.~\ref{ITLS}.
In Fig.~\ref{Imap}(a), $I$ is plotted as a function of the coupling $g$ and the cavity linewidth $\kappa$ while $\gamma$ and $\gamma^*$ are fixed.
Two regions of high indistinguishability are found, which are discussed below.
The one in the upper-right corner corresponds to very large couplings $g$ and broad cavities such as $g > \gamma^*$ and $\kappa >\gamma^*$, which is extremely challenging to reach experimentally for strongly dissipative emitters (i.e. for large values of $\gamma^*$).
%corresponds to the conventional strategy where Purcell effect is exploited until the strong-coupling regime is reached \cite{}.
% Note that for higher ($g$,$\kappa$) values, well beyond realistic values, near-unitary indistinguishability is calculated. 
The other region of high indistinguishability, in the lower-left corner, appears for good cavities for $\kappa<\gamma$ together with moderate or small coupling values $g$. As these values are within experimental reach, this unconventional regime is highly attractive for the generation of indistinguishable photons.

\begin{figure}
\begin{centering}
\includegraphics[width=0.43\textwidth]{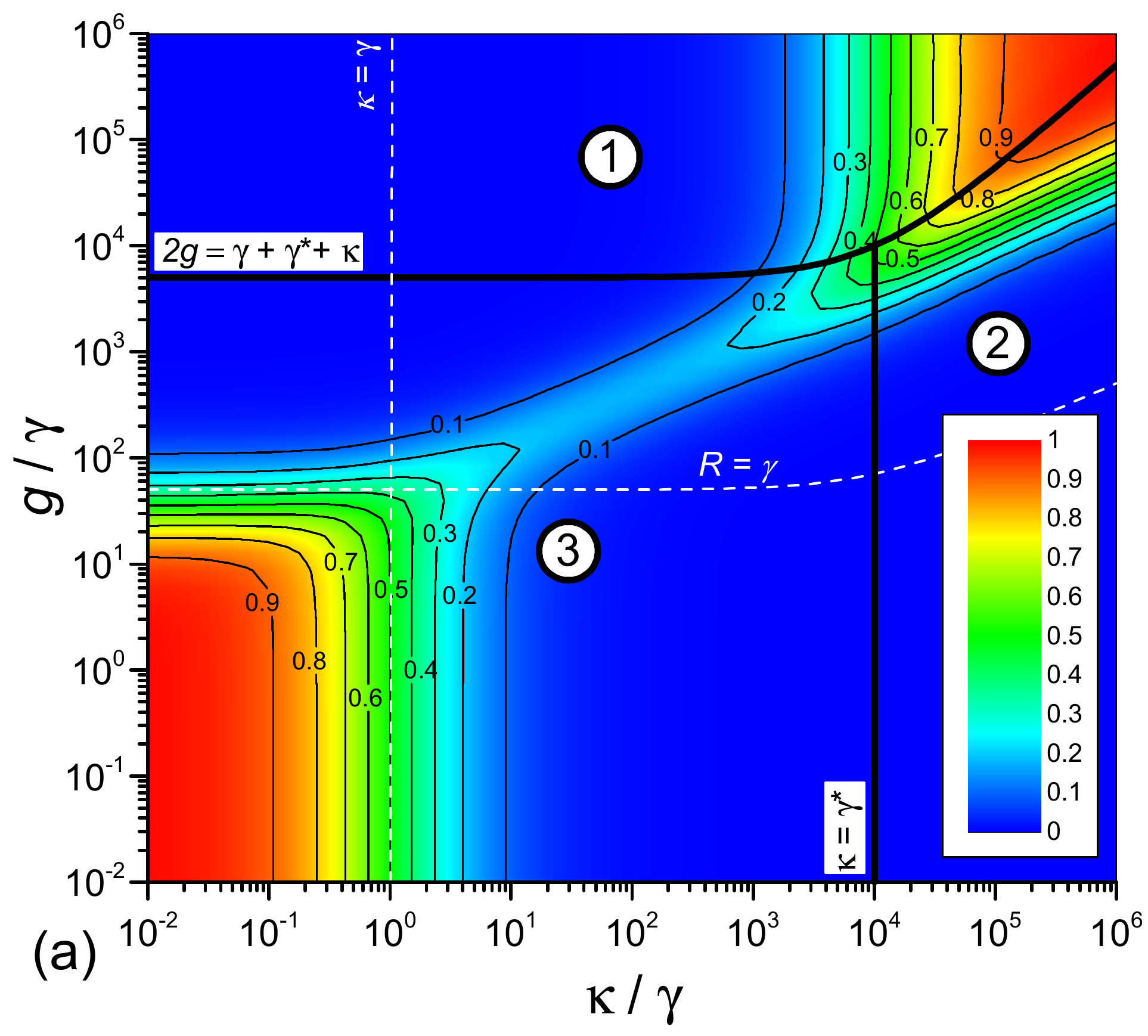}
\includegraphics[width=0.43\textwidth]{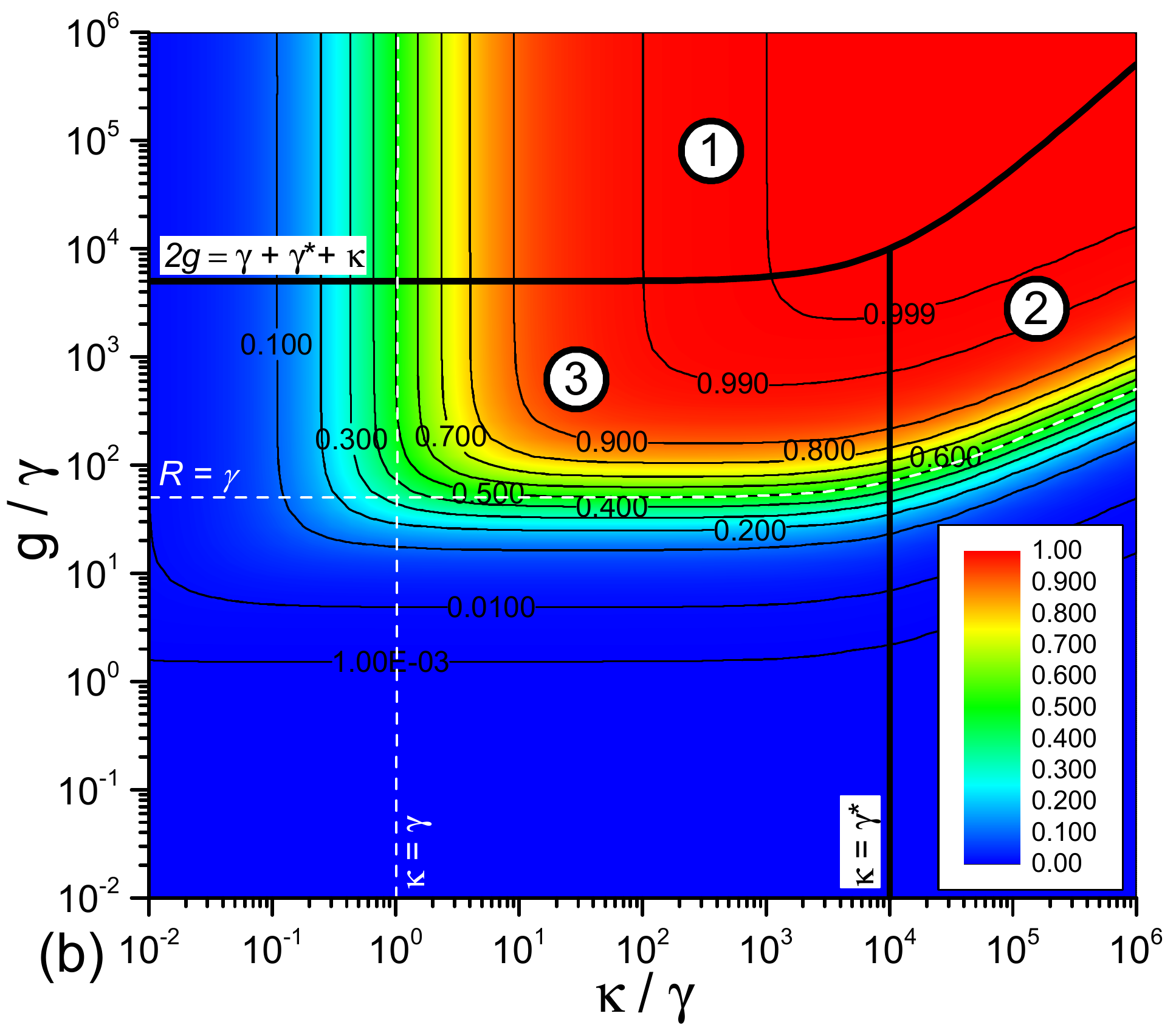}
\includegraphics[width=0.43\textwidth]{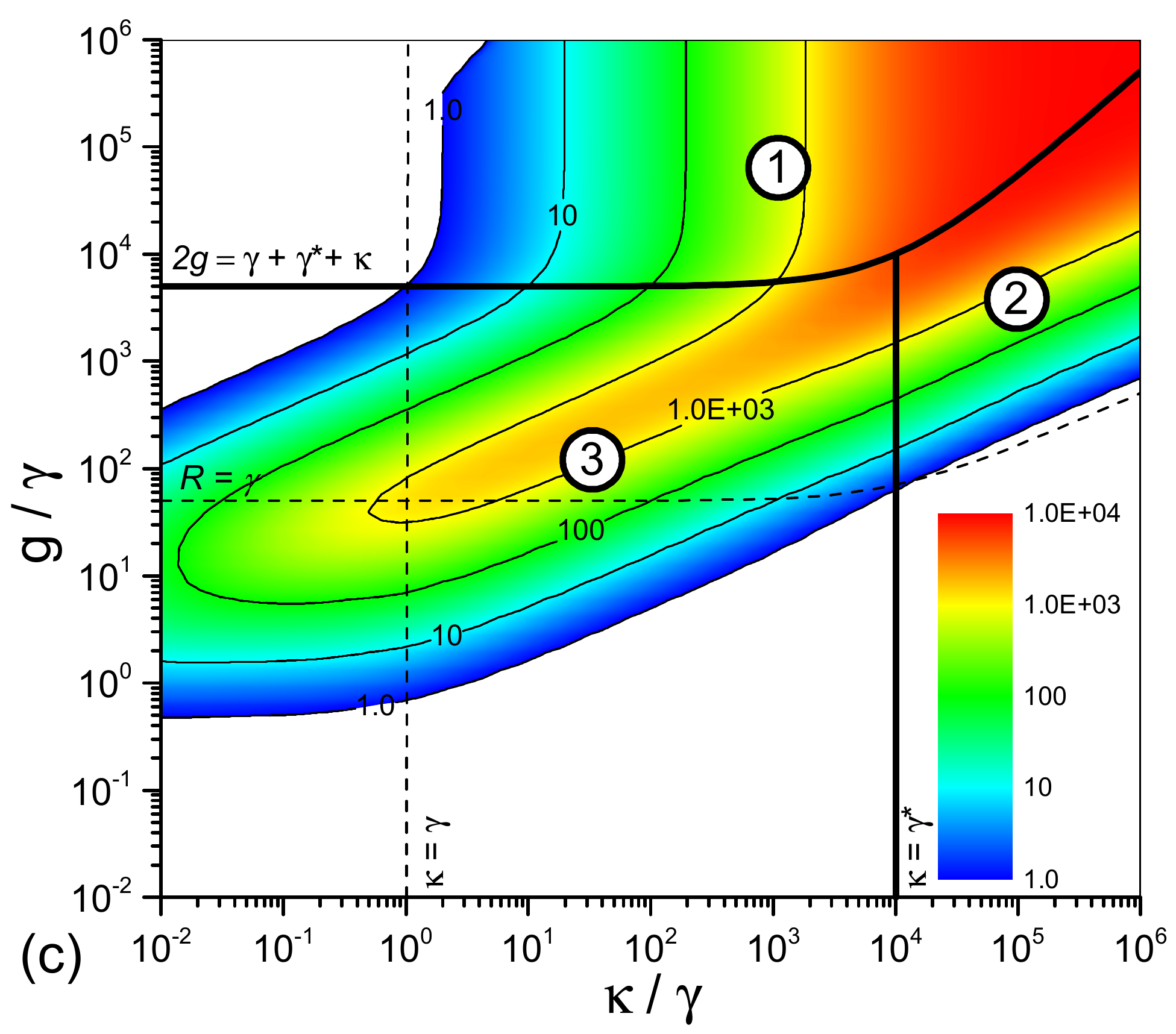}
\par\end{centering}
\caption{(a) Indistinguishability figure of merit ($I$), 
(b) efficiency $\beta$
and (c) funneling ratio $\mathcal{F}$,
plotted as a function of the cavity linewidth $\kappa/\gamma$ and the emitter-cavity coupling strength $g/\gamma$ for a fixed ratio $\gamma^*/\gamma=10^4$.
The funneling ratio is defined as $\mathcal{F} = \beta I \gamma^*/\gamma$.
Solid lines delimit three different regimes discussed in the text: coherent coupling (``1''), incoherent coupling and bad cavity regime (``2''), and incoherent coupling and good cavity regime (``3'').
%The white dashed line corresponds to the limit of cavity funneling where the efficiency-indistinguishability product exceeds the maximum value reachable using spectral filtering on the freely emitted photons.
}
\label{Imap}
\end{figure}

%In order to interpret the result, it is worth to point out that a necessary condition for high indistinguishability is that the photon are Fourier limited \cite{}, i.e. that coherence (field correlations) should extend over the whole photon wavepacket.
%For a sole QE, Eq.\ref{ITLS} means that the linewidth of the photons has to match its decay-induced broadening, i.e. pure dephasing rates should be small compared to recombination rates. For the coupled QE-cavity system studied here, 

%Two conditions (i) Fourier-limited: the width of the mode has to be limited by its (radiative, or at least outside the system) decay. (ii) no time-jitter: the time needed to prepare the mode has to be shorter than its lifetime. or time-jitter = time during which the mode is fed.

%It might be interesting to study different limiting regimes, in which the emission of photons in the cavity can be attributed to a given state of the QE-cavity coupled system.
%To get insigths into these results, we study below the various limiting cases.
To get a physical insight into the calculated indistinguishability, we divide the ($\kappa$,$g$) space into three regimes labelled from ``1'' to ``3'' in Fig.~\ref{Imap}
and study the corresponding limiting cases below. % where the mechanisms limiting the perfect indistinguishability can be isolated and where simple analytic formulae for the indistinguishability degree can be derived.
%where photon emission can be clearly attributed to a given state: either the excited state of the QE, the cavity mode, or the polariton modes.
%In each of this limiting case, the indistinguishability is shown to be determined by the ratio of population decay rate over its total dephasing rate.
First, we can distinguish between the QE-cavity coherent regime occuring for $2g>\kappa + \gamma + \gamma^*$ and the incoherent regime $2g<\kappa + \gamma + \gamma^*$.
In the coherent regime, labelled ``1'' in Fig.~\ref{Imap}, the dynamics consists of damped Rabi oscillations which evolves into an incoherent population of the two polariton modes (i.e. mixed QE-cavity state). In the limit where $2g \gg \kappa  + \gamma + \gamma^*$, the indistinguishability degree reads \cite{supplemental}:
\begin{equation}
I_{\text{cc}} = \frac{(\gamma+\kappa)(\gamma + \kappa + \gamma^*/2)}{(\gamma + \kappa + \gamma^*)^2}.
\end{equation}
As this expression is independent of $g$, increasing $g$ alone does not allow to reach arbitrarily large indistinguishability, as seen on Fig.~\ref{Imap}. On the contrary, nearly-perfect indistinguishability occurs in the coherent regime only if $2g \geq \kappa \gg \gamma^*$.
%This means that, in order to reach near-unity indistinguishability in this regime, 
%the indistinguishability scales as 
%$I \sim 1-3\gamma^*/2\kappa$. For example, $\kappa \sim 15 \gamma^*$ is needed to reach $I= 0.9$ in this strong-coupling regime.
%Hence requirement for reaching high indistinguishability in the SCR are technologically very challenging.

%We will restrict the following discussion to weak coupling, and strong coupling is discussed in supplementary materials \cite{}.
%We now turn to the discussion of weak-coupling regime.
In the incoherent limit ($2g \ll \kappa + \gamma + \gamma^*$), the dynamics of the system can be described in terms of incoherent population transfer with an effective transfer rate between the QE and the cavity given by \cite{auffeves2009pure,auffeves2010controlling}:
\begin{equation}
R = \frac{4g^2}{\kappa +\gamma+\gamma^*}.
\end{equation}
Within this incoherent regime, we can further define a bad-cavity regime for $\kappa > \gamma + \gamma^*$ (labelled ``2'' in Fig.~\ref{Imap}) and a good-cavity regime for $\kappa < \gamma + \gamma^*$ (labelled ``3'' in Fig.~\ref{Imap})  \cite{auffeves2008spontaneous,auffeves2009pure}.
In the bad-cavity limit $\kappa \gg \gamma + \gamma^*$, %spontaneous emission is accellerated by a Purcell rate of $R$ into the cavity.
%In this case, it turns out that
the cavity can be adiabatically eliminated, and its sole effect is to add a new channel of irreversible radiative emission at a rate $R$. Reabsorption by the QE is then negligible.
%This transfer rate is mainly irreversible, as $\kappa>R$.
The dynamics of the coupled system can then be described by the one of an effective QE with a decay rate $\gamma + R$. Applying Eq.~\ref{ITLS} to this effective QE leads to an indistinguishability of
\begin{equation}
I_{\text{bc}} = \frac{\gamma+R}{\gamma + R + \gamma^*} .
\label{Ibc}
\end{equation}
Within this regime of incoherently-coupled bad cavity, the usual strategy to increase indistinguishability is basically to maximize $R$. This can be done by increasing $g$ and/or minimizing $\kappa$.
%However, it ends up when strong-coupling regime is reached as discussed above.
However, from Eq.~\ref{Ibc}, near-unity indistinguishability requires $R\gg\gamma^*$ and consequently $2g \gg \gamma^*$.% and $\kappa \gg \gamma^*$.
%, which ends up when $2g$ reaches $\kappa$ or if $\kappa$ is decreased below $\gamma^*$.
%As a consequence, high indistinguishable values in this regime only occur in a limited region at the vinicity of the coherent coupling regime.
It is found that the coupling strength $g$ has to exceed $\gamma^*$ by nearly one order of magnitude in order to reach an indistinguishability value of $I=0.9$. Reaching such coupling is technologically extremely challenging for solid-state emitters under ambient temperature.
%This regime of weakly-coupled bad cavities regime is well-known in the litterature \cite{}, in which an increase of $g$ (and consequently $R$) results in an increase of indistinguishability. 

On the other hand, the incoherent good-cavity regime (labelled ``3'' in Fig.~\ref{Imap}) occurs for $\kappa < \gamma + \gamma^*$. In this regime, the cavity can store the photons within a time scale comparable to or longer than the QE dephasing time.
The cavity itself then acts as an effective emitter incoherently pumped by the QE \cite{supplemental}, so that the cavity field correlations read
\begin{equation}
\langle a^{\dagger}(t+\tau)a(t)\rangle\ = \rho_{cc}(t)e^{-\Gamma_c \tau/2},
\label{CorrGC}
\end{equation}
  where $\rho_{cc}(t)$ is the population of the cavity mode and $\Gamma_c$ is the linewidth of the cavity-like eigenstate.
From Eq.~\ref{retardedGF}, one can derive that $\Gamma_c = \kappa+R$, which is the sum of the cavity decay rate $\kappa$ into EM modes plus the incoherent reabsorption rate $R$ between the cavity and the QE.
By solving the population rate equations and by plugging in the resulting cavity dynamics $\rho_{cc}(t)$ in Eqs.~\ref{CorrGC} and Eq.~\ref{Iformula}, one finds an indistinguishability of \cite{supplemental}:
\begin{equation}
I_{\text{gc}} = \frac{\gamma + \frac{\kappa R}{\kappa + R}}{\gamma + \kappa + 2R}
\label{Igc}
\end{equation}
Consequently, large indistinguishability occurs in this regime for $\kappa<\gamma$ and $R<\gamma$ (i.e. $g<\sqrt{\gamma \gamma^*}/2$) , in agreement with the full calculation shown on Fig.~\ref{Imap}(a).
This can be understood by noting that two ingredients are involved in the degradation of indistinguishability in this good-cavity regime where the cavity acts as the effective emitter.
The first point is that the initial incoherent feeding of the cavity occurs on a time scale $1/\gamma$, producing a time uncertainty in the population of the cavity. Hence $\kappa$ has to be kept small compared to $\gamma$ in order to prevent such time-jittering effect, in analogy to the incoherent pumping of QE via high energy states \cite{kiraz2004quantum}.
The second point is that, after the initial filling of the cavity, incoherent exchange processes between the QE and the cavity can still occur.
However, back and forth incoherent hopping between the cavity and the QE leads to the dephasing of the photons emitted by the cavity.
To prevent such detrimental hopping, $R<\gamma$ is required.

%\begin{figure}
%\begin{centering}
%\includegraphics[width=0.48\textwidth]{Beta}
%\par\end{centering}
%\caption{Efficency $\beta$ plotted as a function of the cavity linewidth $\kappa$ and the emitter-cavity coupling strength $g$.}
%\label{Betamap}
%\end{figure}

We now discuss the efficiency of the photon emission from the cavity mode, i.e. the probability to have emission by the cavity mode per initial excitation of the QE. The efficiency of photon emission in the cavity mode is given by
\begin{equation}
\beta= \kappa \int_{0}^{\infty}  \langle a^{\dagger}(t)a(t)\rangle.
\label{Betaformula}
\end{equation}
In Fig.~\ref{Imap}(b) the efficiency $\beta$ is plotted as a function of the cavity linewidth $\kappa$ and the emitter-cavity coupling strength $g$.
Near-unitary (i.e. on-demand) efficiencies are obtained in the upper-right corner.
In the weak-coupling regime, we find
\begin{equation}
\beta= \frac{\kappa R}{\kappa R + \gamma(\kappa + R)}
\label{BetaWeak}
\end{equation}
Efficiencies larger than 0.5 are typically obtained for $R>\gamma$ and $\kappa>R$.
This is compatible with high indistinguishability in the region of high $g$ and high $\kappa$ values (i.e. right-upper corner in Fig.~\ref{Imap}), but not in the good cavity regime (i.e. region ``3'' in Fig.~\ref{Imap}).
%Consequently the nearly-perfect indistinguishability region in the good-cavity regime is not compatible with on-demand behavior of the single photon source.
Nevertheless, as discussed in the following, the product of efficiency and indistinguishability $\beta I$ in the good cavity regime can still be way above the one obtained by any linear spectral filtering technique.
%of an equivalent device, where the cavity would only play the role of a ``built-in" spectral filter.
%\begin{figure}
%\begin{centering}
%\includegraphics[width=0.48\textwidth]{BetaIb}
%\par\end{centering}
%\caption{Funneling ratio defined as $\mathcal{F} = \beta I \gamma^*/\gamma$ plotted as a function of the cavity linewidth $\kappa$ and the emitter-cavity coupling strength $g$.}
%\label{BetaImap}
%\end{figure}
Let us consider a linear spectral filter, with a narrow spectral range $\Delta\nu_f$, through which the spectrum of the broad QE is sent. We assume $\Delta\nu_f \ll \gamma^*$. The output efficiency is bounded by $\beta_f \leq \Delta\nu_f / \gamma^*$. Due to the Fourier-transform condition, the corresponding indistinguishability is bounded by
$I_f \leq \gamma / \Delta\nu_f$.
%is given by $I_f = \gamma / \max(\gamma,\Delta\nu) $
Hence the efficiency-indistinguishability product for spectral filtering cannot exceed
$
\beta_f I_f \leq  \gamma / \gamma^*
$.
%This regime, that we denote the ``cavity-funneling'' regime, is discussed in the following.
In order to compare $\beta \times I$ in the present cavity-QED scheme with the upper limit for spectral filtering, we define a cavity-funneling factor
%the product $\beta \times I$  with the maximum reachable value with a posteriori spectral filtering in the absence of a cavity.
%We can define a funneling ratio
\begin{equation}
\mathcal{F} =   \frac{\gamma^*}{\gamma} \beta I,
\end{equation}
such that $\mathcal{F}$ values larger than unity necessarly indicate a spectral cavity-funneling effect.
%More generally
$\mathcal{F}$ indicates the minimum enhancement ratio of $\beta \times I$ with respect to any spectral-filtering effect. In practice this enhancement will be larger since light emitted from a cavity can be very efficiently collected \cite{gazzano2013bright}, in contrast to free-space spontaneous emission.
%Note that the funneling ratio described here is a purely spectral effect and does not describe the further geometrical enhancement of the photon collection efficiency in presence of cavity with respect to free spontaneous emission \cite{gazzano2013bright}.
In Fig.~\ref{Imap}(c), the funneling $\mathcal{F}$ is plotted in the same parameter range ($\kappa$,$g$) as previous plots. Only the values satisfying the cavity-funneling condition of $\mathcal{F}>1$ are shown.
%On Fig.~\ref{BetaImap} only the values verifying the necessary condition for cavity funneling $\beta I \geq  \gamma /\gamma^*$ are shown.
%(This region is also indicated by a dashed-line contour on the indistinguishability map (Fig.~\ref{Imap}).)
It appears clearly that almost-perfect indistinguishability in the good cavity regime is compatible with cavity funneling.
In Fig.~\ref{Comp}, $I$, $\beta$ and $\mathcal{F}$ are plotted as a function of $\kappa$ for a fixed value of $g$. The full calculation is found to be in good agreement with the above formulae for the incoherent regime. It illustrates the necessary trade-off between indistinguishability and efficiency in the good-cavity regime, where a clear maximum of the funneling factor occurs.
%Owing to the Fourier transform condition,
The large calculated values for $\mathcal{F}$ 
are signatures of a very efficient redirection of the QE spectrum into the unperturbed cavity spectrum of linewidth $\kappa$.
%Indeed, while the QE decay rate is not necessarly affected, the Fourier transform condition indicates that 
%In other words, 
%the emission rate at the cavity frequency is speed-up by at least
%$\mathcal{F}$ 
%with respect to free-space emission.

%. The enhancement of the  and will be in practice multiply by geometrical enhancement of the photon collection efficiency due to the cavity \cite{gazzano2013bright}.

\begin{figure}
\begin{centering}
\includegraphics[width=0.48\textwidth]{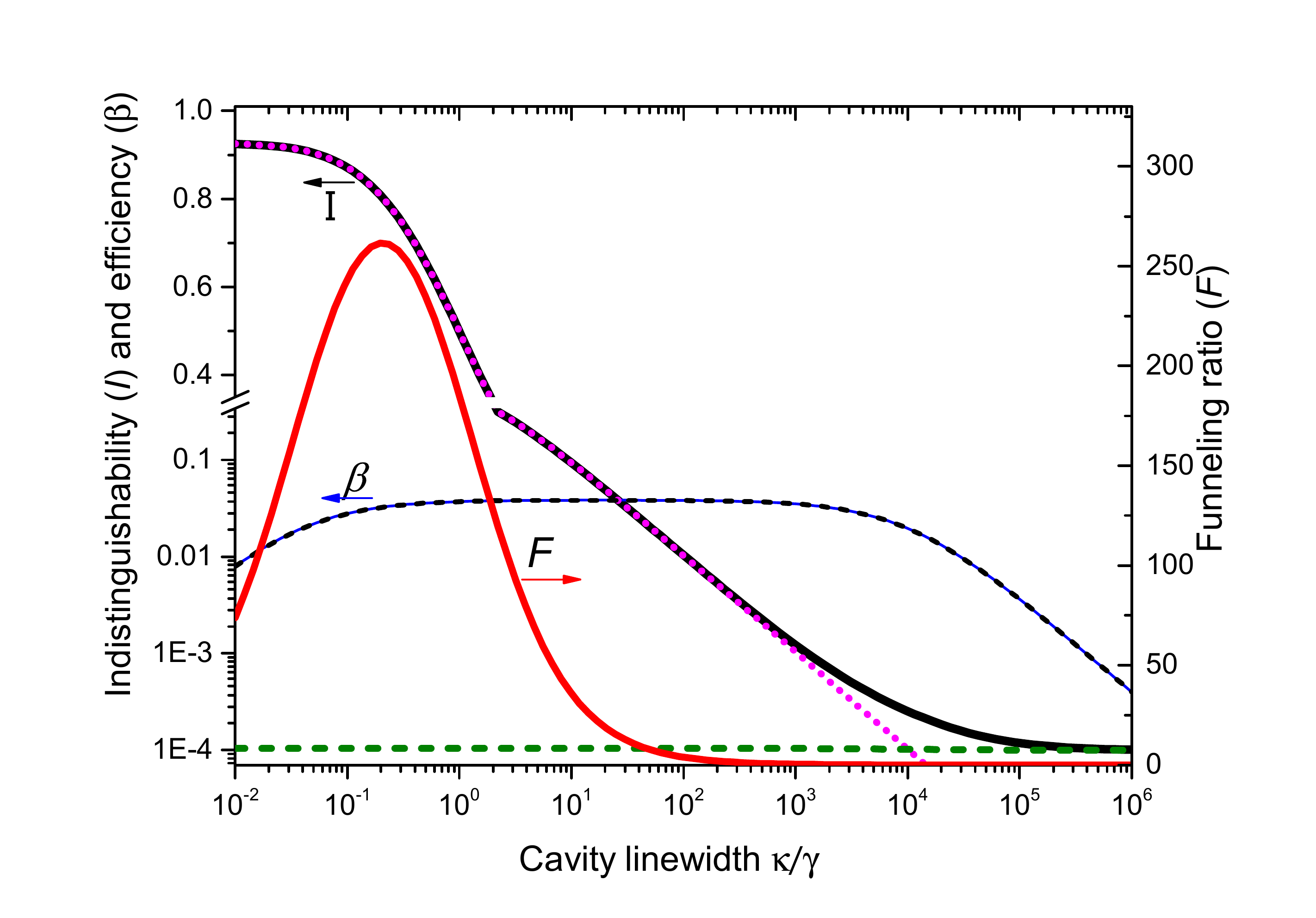}
\par\end{centering}
\caption{Indistinguishability figure of merit $I$ (full calculation in black solid line), efficiency $\beta$ (blue solid line) and cavity funneling ratio $\mathcal{F}$ (red solid line) for a fixed emitter-cavity coupling of $g = 10 \gamma$. The full calculation for indistinguishability is compared with analytic expressions in two different limiting cases: incoherently-coupled good cavity regime (dotted line, Eq.~\ref{Igc}) and coherently-coupled bad cavity regime (dashed line, Eq.~\ref{Ibc}). The analytical expression for $\beta$ (dotted line, Eq.~\ref{BetaWeak}) overlaps perfectly with the full calculation.}
\label{Comp}
\end{figure}
 
%\begin{figure}
%\begin{centering}
%\includegraphics[width=0.15\textwidth]{LambdaSchem}
%\par\end{centering}
%\caption{Schematic of the $\Lambda$-scheme where heralding of the photons emitted in the cavity is proposed.}
%\label{LambdaSchem}
%\end{figure}
 
%To illustrate the feasibility of reaching this unconventional regime at room temperature, we propose two possible experimental realizations.
Finally, we propose two experimental realizations of this unconventional regime at room temperature.
We first consider a single self-assembled quantum dot coupled to a photonic crystal cavity. State of the art photonic crystal cavities can provide $\hbar g$ =120 $\mu$eV and $\hbar \kappa$=20 $\mu$eV \cite{arakawa2012cavity}. Assuming $\hbar\gamma$ = 60 $\mu$eV and $\hbar\gamma^*$ = 7 meV for an InAs/GaAs QD at 300K \cite{heitz1999temperature,borri01},
%\footnote{qdgf\cite{heitz1999temperature}}
we predict $I$=0.72, $\beta$=0.088 and $\mathcal{F}$=7.3.
Secondly, we consider a single silicon vacancy (SiV) center in a nano-diamond coupled to a fiber cavity. For SiV at 300K, we take $\gamma$ = 2$\pi$$\times$ 160 MHz and $\gamma^*$ = 2$\pi$$\times$ 550 GHz \cite{neu2011single}.
Coupling SiV with a fiber cavity with $g$=2$\pi$$\times$1.0 GHz and $\kappa$=2$\pi$$\times$30 MHz is within experimental reach \cite{kaupp2013scaling}, for which our calculation predicts $I$=0.81, $\beta$=0.035 and $\mathcal{F}$=99.
These predicted degrees of indistinguishability at room temperature are comparable with state of the art values obtained from low temperature single-photon sources under incoherent pumping \cite{gazzano2013bright}, with efficiencies far beyond any spectral filtering technique.

In summary, for strongly dissipative emitters we predict an unconventional regime of high indistinguishability
%using low cavity linewidth (i.e. $\kappa<\gamma$) and relatively modest coupling strength, where
in which the broad spectrum of the quantum emitter is funneled into a narrow cavity resonance. %, in which cavity funneling occurs.
For typical room-temperature quantum emitters, the associated efficiency can surpass any spectral filtering schemes by orders of magnitude.
%In addition, an heralding scheme of the cavity emission is proposed based on $\Lambda$-like quantum structure.
This strategy opens the road towards the generation of indistinguishable single photons from solid-state quantum emitters under ambient temperature. %with an efficiency exceeding the spectral filtering one.

This work has been funded by the European Union's Seventh Framework Programme (FP7) under Grant Agreement No. 618078 (WASPS). Jason Smith and John Rarity are acknowledged for fruitfuil discussions.

\clearpage

\setcounter{equation}{0}
\setcounter{figure}{0}
\setcounter{table}{0}
\setcounter{page}{1}
\makeatletter
\renewcommand{\theequation}{S\arabic{equation}}
\renewcommand{\thefigure}{S\arabic{figure}}
\renewcommand{\bibnumfmt}[1]{[S#1]}
\renewcommand{\citenumfont}[1]{S#1}

\title{Supplemental Material: Cavity-funneled generation of indistinguishable single photons from strongly dissipative quantum emitters}

\author{Thomas Grange}
\affiliation{Universit\'e Grenoble-Alpes, ``Nanophysics and Semiconductors'' joint team, 38000 Grenoble, France}
\affiliation{CNRS, Institut N\'{e}el, ``Nanophysics and Semiconductors'' joint team, 38000 Grenoble, France}

\author{ Gaston Hornecker}
\affiliation{Universit\'e Grenoble-Alpes, ``Nanophysics and Semiconductors'' joint team, 38000 Grenoble, France}
\affiliation{CNRS, Institut N\'{e}el, ``Nanophysics and Semiconductors'' joint team, 38000 Grenoble, France}
\affiliation{CEA, INAC-SP2M, ``Nanophysics and Semiconductors'' joint team, 38000 Grenoble, France}

\author{David Hunger}
\affiliation{Fakult\"{a}t f\"ur Physik, Ludwig-Maximilians-Universit\"at, Schellingstr. 4, 80799 M\"unchen, Germany}
 
  \author{Jean-Philippe Poizat}
\affiliation{Universit\'e Grenoble-Alpes, ``Nanophysics and Semiconductors'' joint team, 38000 Grenoble, France}
\affiliation{CNRS, Institut N\'{e}el, ``Nanophysics and Semiconductors'' joint team, 38000 Grenoble, France}
   
   \author{Jean-Michel G\'erard}
 \affiliation{Universit\'e Grenoble-Alpes, ``Nanophysics and Semiconductors'' joint team, 38000 Grenoble, France}
\affiliation{CEA, INAC-SP2M, ``Nanophysics and Semiconductors'' joint team, 38000 Grenoble, France}

 \author{Pascale Senellart}
 \affiliation{CNRS, Laboratoire de Photonique et de Nanostructures,  91460 Marcoussis, France}
  
 \author{Alexia Auff\`{e}ves}
\affiliation{Universit\'e Grenoble-Alpes, ``Nanophysics and Semiconductors'' joint team, 38000 Grenoble, France}
\affiliation{CNRS, Institut N\'{e}el, ``Nanophysics and Semiconductors'' joint team, 38000 Grenoble, France}

\date{\today}

\maketitle

\begin{widetext}
\vspace{1cm}
\begin{center}
\textbf{\large Supplemental Materials}
\end{center}
\end{widetext}

\section{Hamiltonian and master equation}

The Hamiltonian of a two-level quantum emitter (QE) interacting with a quantized mode of an optical cavity can be written as:
\begin{equation}
\hat{H}= \hbar \omega_{\text{QE}} \hat{c}^{\dagger}\hat{c} + \hbar \omega_{\text{cav}} \hat{c}^{\dagger}\hat{c} + \hbar g(\hat{c}+\hat{c}^{\dagger})(\hat{a}+\hat{a}^{\dagger}),
\end{equation}
where $\hat{c}^{\dagger}$ and $\hat{c}$ are fermionic creation and annihilation operators for the QE while $\hat{a}^{\dagger}$ and $\hat{a}$ are bosonic creation and annihilation operators for the cavity. $\omega_{\text{QE}}$ is the QE frequency, $\omega_{\text{cav}}$ is the cavity frequency, and $g$ is the QE-cavity coupling strength.
We consider up to one excitation in the system so that within the rotating-wave approximation the dynamics involved only the states $\{|g,0\rangle,|e,0\rangle,|g,1\rangle\}$.
As no coherent coupling occurs between $|g,0\rangle$ and the other states, it is sufficient to study the dynamics within the basis
formed by
$\{|e,0\rangle,|g,1\rangle\}$, in which the Hamiltonian reads

\begin{equation}
\hat{H}= \begin{pmatrix}
0& g\\
  g & \delta
 \end{pmatrix} ,
\end{equation}
where $\delta = \omega_{\text{cav}} -\omega_{\text{Qe}}$ is the detuning between the QE and the cavity.
The density matrix can be written as
\begin{equation}
\rho(t) = \begin{pmatrix}
  \rho_{ee}(t) = \langle \hat{c}^{\dagger} (t) \hat{c}(t) \rangle & \rho_{ec}(t)=\langle \hat{c}^{\dagger} (t) \hat{a}(t) \rangle \\
\rho_{ce}(t)  = \langle \hat{a}^{\dagger} (t) \hat{c}(t) \rangle & \rho_{cc}(t) = \langle \hat{a}^{\dagger} (t) \hat{a}(t) \rangle
 \end{pmatrix}
\end{equation}

Its evolution is described by the following Linbladt master equation \cite{petruccione2002theory,auffeves2009pure}:
\begin{equation}
\frac{\partial \hat{\rho}}{\partial t} = \mathcal{L} [\rho]
\label{master}
\end{equation}

\begin{equation}
\hat{\mathcal{L}} [\rho] = \frac{i}{\hbar}[\hat{\rho},\hat{H}] + \hat{\mathcal{L}}_{\text{QE}} + \hat{\mathcal{L}}_{\text{cav}} + \hat{\mathcal{L}}_{\text{deph}}
\end{equation}

where the dissipative terms describing respectively the QE decay, the cavity decay and pure dephasing reads in the $\{|e,0\rangle,|g,1\rangle\}$ basis

\begin{equation}
\hat{\mathcal{L}}_{\text{QE}}[\hat{\rho}] =
-\gamma
\begin{pmatrix}
  \rho_{ee}&  \rho_{ec}/2\\
  \rho_{ce}/2 & 0
 \end{pmatrix}
\end{equation}

\begin{equation}
\hat{\mathcal{L}}_{\text{cav}}[\hat{\rho}] =
-\kappa 
\begin{pmatrix}
0 & \rho_{ec}/2\\
   \rho_{ce}/2 & \rho_{cc}
 \end{pmatrix}
\end{equation}

\begin{equation}
\hat{\mathcal{L}}_{\text{deph}}[\hat{\rho}] =
 -\gamma^* 
\begin{pmatrix}
0 &\rho_{ec}/2 \\
 \rho_{ce}/2  &  0
 \end{pmatrix}
\end{equation}

An initial excitation of the QE is assumed, i.e. $\hat{\rho}(0)=|e,0\rangle \langle e,0 |$. The evolution of the density matrix is then computed using
\begin{equation}
\hat{\rho}(t) = e^{\mathcal{L}t} |e,0\rangle \langle e,0 |
\end{equation}

%Equivalently, these Markovian dissipative terms are described within the NEGF formalism by
%The two-time correlators can be obtained via the quantum regression theorem.

\section{Calculation of two-time correlators}

In order to conveniently express the two-time correlators, we make use of the non-equilibrium Green's function formalism \cite{haug2008quantum}.
Here the lesser, greater and retarded Green's function are respectively defined by:

\begin{equation}
\hat{G}^<(t+\tau,t) = \begin{pmatrix}
  \langle \hat{c}^{\dagger} (t+\tau) \hat{c}(t) \rangle & \langle \hat{c}^{\dagger} (t+\tau) \hat{a}(t) \rangle \\
  \langle \hat{a}^{\dagger} (t+\tau) \hat{c}(t) \rangle & \langle \hat{a}^{\dagger} (t+\tau) \hat{a}(t) \rangle
 \end{pmatrix} ,
\end{equation}

\begin{equation}
\hat{G}^>(t+\tau,t) = \begin{pmatrix}
  \langle \hat{c}(t)\hat{c}^+ (t+\tau)  \rangle & \langle \hat{a}(t)\hat{c}^+ (t+\tau)  \rangle \\
  \langle \hat{c}(t)\hat{a}^+ (t+\tau)  \rangle & \langle \hat{a}(t)\hat{a}^+ (t+\tau)  \rangle
 \end{pmatrix} ,
\end{equation}

\begin{equation}
\hat{G}^{R}(\tau) = \Theta(\tau)\left[ G^>(t+\tau,t)+G^<(t+\tau,t) \right] ,
\end{equation}
where $\tau$ is a time difference.
Note that for simplicity we have dropped in the above definitions the factor $i$ involved in standard definitions  \cite{haug2008quantum} .
The retarded and lesser self-energies describing the dissipative terms are expressed as:

\begin{equation}
\hat{\Sigma}^R(t+\tau,t) = 
\delta(\tau)
\begin{pmatrix}
  (\gamma+\gamma^*)/2 & 0 \\
  0 & \kappa/2
 \end{pmatrix}
\end{equation}

\begin{equation}
\hat{\Sigma}^<(t+\tau,t) = 
\delta(\tau)
\begin{pmatrix}
  \gamma^* G^<_{e,e}(t+\tau,t) & 0 \\
  0 & 0
 \end{pmatrix}
\end{equation}

For such time-independent Hamiltonian the retarded Green's function depends only of one variable. It can be expressed in angular frequencies as:
\begin{equation}
\hat{G}^{R}(\omega) = -i \int d \omega e^{i\omega t} \hat{G}^R(\tau).
\end{equation}
%where the imaginary part allows the definition of $\hat{G}^{R}(\omega)$ to match the standard one.
The Dyson's equation for the retarded Green's function reads:

\begin{equation}
\hat{G}^{R}(\omega) = (\omega - H - \Sigma^R(\omega))^{-1}
\end{equation}

\begin{equation}
\hat{G}^{R}(\omega) = \begin{pmatrix}
  \omega + i\gamma/2 + i\gamma^*/2 & g \\
  g & \omega -\delta + i\kappa/2
 \end{pmatrix}^{-1}
\label{gr}
\end{equation}

%The lesser Green's function at equal time corresponds to the density matrix.
%One can easily check that Kadanoff-Baym equations including these self-energies leads to the same master equation for the density matrix.

From Kadanoff-Baym equations, which describe the equations of motion of the Green's functions \cite{haug2008quantum}, one can easily derive the following relation in the case of Markovian self-energies for $\tau>0$:
\begin{equation}
\hat{G}^<(t+\tau,t) = \hat{G}^R(\tau) \hat{G}^<(t,t) = \hat{G}^R(\tau) \hat{\rho} (t).
\end{equation}

From this relation we can extract the cavity correlations
\begin{equation}
\begin{split}
\langle \hat{a}^{\dagger}(t+\tau)\hat{a}(t) \rangle & = \langle g,1 |  \hat{G}^R(\tau) \hat{\rho} (t) |g,1\rangle \\
& = G^R_{cc}(\tau) \rho_{cc} (t) + G^R_{ce}(\tau) \rho_{ec} (t)
\end{split}
\label{cc}
\end{equation}

Hence the calculation of the two-time correlators is splitted into the calculation of two one-time operators, namely the density matrix  $\hat{\rho} (t) $ on one hand and the retarded Green's function $\hat{G}^R(\tau)$ on the other hand. $\hat{G}^R(\tau)$ can be computed by Fourier transforming Eq.~\ref{gr}, or by solving its equation of motion which reads
\begin{equation}
i \frac{\partial}{\partial \tau} \hat{G}^R(\tau) = i \delta(\tau)\hat{\mathrm{I}} + \left[\hat{H} -  i \hat{\Sigma}^R(0)\right] \hat{G}^R(\tau) .
\label{eomGR}
\end{equation}

\section{Indistinguishability in limiting cases of cavity-QED}

In the following discuss three limiting cases of dissipative cavity quantum electrodynamics (cavity-QED) and derive the degree of photon indistinguishability in each of these cases. We assume a perfect resonance (i. e. $\delta=0$) between the QE and the cavity.

\subsection{Coherent coupling regime}
Coherent-coupling regime between the QE and the cavity occurs if $2g>\kappa + \gamma + \gamma^* $.
In the limit $2g\gg \kappa + \gamma + \gamma^* $, we derive below an analytical expression for the indistinguishability. % in a qualitative way.
%To do so we separate between two contributions, namely (i) the photons emitted during the initial Rabi oscillations and (ii) the one emitted after thermalization between the two polariton mixed states.
%During the initial coherent oscillations (i.e. Rabi oscillations), the mean decay rate is $(\gamma+\kappa)/2$, while the mean dephasing rate reads $(\kappa+\gamma+\gamma^*)/2$. The corresponding  indistinguishability is given by $(\gamma+\kappa)/(\kappa + \gamma+\gamma^*)$ according to Eq.~2 of the paper. The proportion of photon emitted during this coherent evolution is given by $(\gamma+\kappa)/(\kappa+ \gamma + \gamma^*)$.
%These Rabi oscillations are followed by a regime of incoherent population of the two polariton modes, where each polariton decays as $(\gamma+\kappa)/2$, with a pure dephasing rate of $\gamma^*/2$. The corresponding indistinguishability is given by $(1/2)\times (\gamma+\kappa)/(\gamma + \kappa + \gamma^*)$, where the factor $1/2$ origins from the fact that these two polariton modes are incoherently populated.
%Summing up these two contributions with their relative weight leads to
In this limit, the coherent part of the dynamics (i.e. Rabi oscillations) is much faster than the incoherent part (i.e. population and phase decay). An approximate solution of the dynamics is obtained by decoupling these two timescales. The density matrix then reads:

\begin{widetext}

%\begin{equation}
%\rho(t) \simeq 
%e^{-(\gamma+\kappa)t/2} 
%\begin{pmatrix}
%\cos^2(gt) e^{-\gamma^* t/2}   +\frac{1-e^{-\gamma^* t/2}}{2} & e^{-\gamma^* t/2}\frac{-\sin(2gt)}{2i}  \\
%e^{-\gamma^* t/2}\frac{\sin(2gt)}{2i} & \sin^2(gt) e^{-\gamma^* t/2} +\frac{1-e^{-\gamma^* t/2}}{2}
% \end{pmatrix} 
%\end{equation}

\begin{equation}
\rho(t) \simeq 
e^{-(\gamma+\kappa)t/2}
\left[
e^{-\gamma^* t/2}
\begin{pmatrix}
\cos^2(gt)    & \frac{-\sin(2gt)}{2i}  \\
\frac{\sin(2gt)}{2i}  & \sin^2(gt) 
 \end{pmatrix} 
 +
(1-e^{-\gamma^* t/2})
 \begin{pmatrix}
\frac{1}{2}    & 0  \\
0  & \frac{1}{2}
 \end{pmatrix} 
  \right] ,
\end{equation}
where the first term describes the initial damped Rabi oscillations, while the second term accounts for the incoherent part of the dynamics after dephasing.
The retarded Green's function reads

\begin{equation}
G^R(\tau) \simeq 
e^{-(\gamma+\kappa+\gamma^*)\tau/4} 
\times
\begin{pmatrix}
  \cos(g\tau) &  -i\sin(g\tau)  \\
-i\sin(g\tau)   & \cos(g\tau)
 \end{pmatrix}.
\end{equation}

By averaging over the fast coherent dynamics,  the cavity population and the cavity correlations are given respectively by

\begin{equation}
< \rho_{cc}(t) > = \frac{1}{2} e^{-(\gamma+\kappa)t/2} ,
\end{equation}

\begin{equation}
<\vert  \langle \hat{a}^{\dagger}(t+\tau) \hat{a}(t) \rangle \vert^2> %= <\vert G^R_{cc}(\tau) \vert^2 > < \vert \rho_{cc} (t) \vert^2>\\
%&  + <\vert G^R_{ce}(\tau) \vert^2 > < \vert \rho_{ec} (t)\vert^2> \\
 = \frac{1}{4} \left [
  e^{-(\gamma+\kappa + \gamma^*)t}
 +e^{-(\gamma+\kappa)t}
\right]
\frac{e^{-(\gamma+\kappa+\gamma^*)\tau/2}}{2}
,
\end{equation}
where the $<>$ indicates an average over the fast rotating terms.
It gives a photon indistinguishability of

\begin{equation}
I_{\text{cc}}  =  \frac{\int_{0}^{\infty} dt  \int_0^{\infty} d\tau  <\vert  \langle \hat{a}^{\dagger}(t+\tau) \hat{a}(t) \rangle \vert^2> }
{\frac{1}{2} \vert \int_{0}^{\infty} dt < \rho_{cc}(t) > \vert^2}
= \frac{(\gamma+\kappa)(\gamma + \kappa + \gamma^*/2)}{(\gamma + \kappa + \gamma^*)^2}.
\end{equation}

\end{widetext}

\subsection{Incoherent-coupling regime}

The incoherent limit occurs for $2g \ll \kappa + \gamma + \gamma^*$, for which it is shown below that the coherences can be adiabatically eliminated.

\subsubsection{Adiabatic elimination of coherences}

From the master equation we have at resonance ($\delta=0$):
\begin{subequations}
\begin{equation}
\frac{\partial \rho_{ee}}{\partial t}  = -\gamma \rho_{ee} + ig(\rho_{ec} - \rho_{ce})
\end{equation}
\begin{equation}
\frac{\partial \rho_{cc}}{\partial t}  = -\kappa \rho_{cc} + ig(\rho_{ce}-\rho_{ec} )
\end{equation}
\begin{equation}
\frac{\partial \rho_{ec}}{\partial t}  = -\frac{\gamma+\gamma^*+\kappa}{2}  \rho_{ec} + ig(\rho_{ee} - \rho_{cc})
\label{cohM}
\end{equation}
\end{subequations}
If $\gamma + \gamma^*+\kappa \gg 2g$, 
%adiabatic elimination for the coherences leads to
coherences  can  be adiabatically eliminated in Eq.~\ref{cohM} by setting $\partial \rho_{ec}/\partial t \sim 0 $,  leading to:
%($\partial \rho_{ec}/\partial t \sim 0$ in Eq.~\ref{cohM}):
\begin{equation}
\rho_{ec}(t) \simeq \frac{2ig(\rho_{ee}(t) - \rho_{cc}(t))}{\gamma+\gamma^*+\kappa}.
\end{equation}
We are then left with the following rate equations for the populations
\begin{subequations}
\begin{equation}
\frac{\partial \rho_{ee}}{\partial t}  = -(\gamma+R) \rho_{ee}  + R \rho_{cc},
\end{equation}
\begin{equation}
\frac{\partial \rho_{cc}}{\partial t}  = -(\kappa+R) \rho_{cc} + R \rho_{ee},
\end{equation}
\label{coupled}
\end{subequations}
where the exchange rate between the QE and the cavity reads \cite{auffeves2009pure}
\begin{equation}
R = \frac{4g^2}{\gamma+\gamma^*+\kappa}.
\end{equation}
By solving the coupled rate equations Eq.~\ref{coupled}, we find that the efficiency is given by
\begin{equation}
\beta = \kappa \int dt \rho_{cc}(t) = \frac{\kappa R}{\kappa R + \gamma ( \kappa + R)}.
\end{equation}

%\subsection{Retarded GF}

%\begin{equation}
%i \frac{\partial}{\partial \tau} \hat{G}^R(\tau) = \delta(\tau)\hat{\mathrm{I}} + (\hat{H} +  \hat{\Sigma}^R(0))\hat{G}^R(\tau)
%\end{equation}

\subsubsection{Regime of incoherent coupling and bad cavity}

In the bad cavity limit (i.e. $\kappa \gg \gamma + \gamma^*$), we can eliminate the cavity population (i.e. $\partial \rho_{cc}/\partial t \sim 0 $) in the above rate equations, leading to
\begin{equation}
\rho_{cc}(t) = \frac{R}{\kappa + R} \rho_{ee}(t)
\label{rhocc}
\end{equation}
%In addition we can split the cavity correlation function into two component:
On the other hand, approximation on the retarded Green's function involved in Eq.~\ref{cc} can also be derived. 
%For $\kappa \gg \gamma + \gamma^*$, 
For $\tau > 1/\kappa$, one can adiabatically eliminate the off-diagonal term $\partial G^R_{ce}/\partial \tau \sim 0 $ in Eq.~\ref{eomGR}:
\begin{equation}
G^R_{ce}(\tau) \simeq -2i\frac{g}{\kappa} G^R_{ee}(\tau), 
\end{equation}
so that the second term in Eq.~\ref{cc} reads for $\tau > 1/\kappa$
\begin{equation}
G^R_{ce}(\tau) \rho_{ec} (t) \simeq \frac{R}{\kappa} G^R_{ee}(\tau) \rho_{ee}(t).
\end{equation}
On the other hand, the first term in Eq.~\ref{cc} is given by
\begin{equation}
G^R_{cc}(\tau) \rho_{cc} (t) \simeq \frac{R}{\kappa} e^{-\kappa \tau/2} \rho_{ee}(t).
\end{equation}
 %Suming up these two contributions in Eq.~\ref{cc} give
 For $\tau > 1/\kappa$, only the former term is not vanishing:
\begin{equation}
\langle \hat{a}^{\dagger}(t+\tau)\hat{a}(t) \rangle \simeq \frac{R}{\kappa}  G^R_{ee}(\tau) \rho_{ee}(t).
\end{equation}
The last expression indicates that in this regime the cavity correlations do indeed follow the QE correlations, with:
\begin{equation}
\rho_{ee}(\tau) \simeq e^{-(\gamma+R)t}
\end{equation}
\begin{equation}
G^R_{ee}(\tau) \simeq e^{-(\gamma+\gamma^*+R)\tau/2}.
\end{equation}
As the integral of the correlations (Eq.~2 of the paper) is dominated by delay times such as $\tau > 1/\kappa$, one finds for the indistinguishability
\begin{equation}
I_{\text{bc}} = \frac{\gamma+R}{\gamma + R + \gamma^*} .
\label{Ibc}
\end{equation}

%which means that the cavity correlations directly reflects the QE correlations.

%\begin{subequations}
%\begin{equation}
%\frac{\partial G^R_{ee}}{\partial t}  = -\frac{\gamma+\gamma^*}{2} \rho_{ee} - g(\rho_{ec} + \rho_{ce})
%\end{equation}
%\begin{equation}
%\frac{\partial G^R_{cc}}{\partial t}  = -\frac{\kappa}{2} \rho_{cc} + g(\rho_{ec} + \rho_{ce})
%\end{equation}
%\begin{equation}
%\frac{\partial G^R_{ec}}{\partial t}  = -\frac{\gamma+\gamma^*+\kappa}{2}  \rho_{ec} + g(\rho_{ee} - \rho_{cc})
%\end{equation}
%\end{subequations}

\subsubsection{Regime of incoherent coupling and good cavity}

In the limit $\gamma + \gamma^* \gg \kappa $, we show below that the cavity correlations are dominated by the terms diagonal in the cavity mode.
In this limit, the projection of the retarded Green's function on the cavity mode reads
\begin{equation}
G^R_{cc}(\tau) \simeq e^{-\Gamma_c \tau /2 }
\label{grccgc}
\end{equation}
where $\Gamma_c$ is the linewidth of the cavity-like eigenstate which can be calculated from from Eq.~\ref{gr}:
\begin{equation}
\Gamma_c = \kappa + \frac{4g^2}{\vert \gamma + \gamma^* - \kappa \vert} \simeq \kappa + R.
\end{equation}
In Eq.~\ref{eomGR}, we can adiabatically eliminate $G^R_{ce}$ for $\tau>1/(\gamma+\gamma^*)$, which gives
\begin{equation}
G^R_{ce}(\tau) \simeq -i\frac{g}{\gamma+\gamma^*} e^{-\Gamma_c\tau/2},
\end{equation}
and thus providing an upper limit for the second term in Eq.~\ref{cc}:
\begin{equation}
G^R_{ce}(\tau) \rho_{ec}(t) \le \frac{R}{\gamma+\gamma^*} e^{-\Gamma_c\tau/2} \rho_{ee}(t)
\label{ceec}
\end{equation}
On the other hand, Eq.~\ref{cohM} provides an upper bound for $\rho_{cc}$:
\begin{equation}
\rho_{cc} \ge \frac{R}{R+\kappa}  \rho_{ee}(t),
\end{equation}
which combined with Eq.~\ref{grccgc} gives
\begin{equation}
G^R_{cc}(\tau) \rho_{cc} \ge \frac{R}{R+\kappa} e^{-\Gamma_c\tau/2}  \rho_{ee}(t).
\label{cccc}
\end{equation}
Eqs.~\ref{ceec} and \ref{cccc} show that in this regime  $G^R_{cc} \rho_{cc}$ dominates over $G^R_{ce} \rho_{ec}$ for $\tau>1/(\gamma+\gamma^*)$:
\begin{equation}
\langle \hat{a}^{\dagger}(t+\tau)\hat{a}(t)\rangle\ \simeq G^R_{cc}(\tau) \rho_{cc} (t).
\end{equation}
We can hence express the indistinguishability as if the cavity acts as an effective emitter:
\begin{equation}
I_{gc} =  \frac{\int_{0}^{\infty} dt \rho_{cc}^2(t) \int_0^{\infty} d\tau   e^{-\Gamma_c \tau} }
{\frac{1}{2} \vert \int_{0}^{\infty} dt \rho_{cc}(t) \vert^2}.
\end{equation}
By solving the rate equations for populations (Eq.~\ref{coupled}), and plugging in $\rho_{cc}(t)$ in the above expression, we find after some algebra
\begin{equation}
I_{\text{gc}} = \frac{\gamma + \frac{\kappa R}{\kappa + R}}{\gamma + \kappa + 2R}.
\label{Igc}
\end{equation}

\end{document}